# Plasma induced surface modification of sapphire and its influence on graphene grown by PECVD

Authors:

Miguel Sinusia Lozano [1], Ignacio Bernat-Montoya [1], Todora Ivanova Angelova [1], Alberto Boscá Mojena [2], Francisco J. Díaz-Fernández [1], Miroslavna Kovylina [1], Alejandro Martínez[1], Elena Pinilla Cienfuegos [1], Víctor J. Gómez * [1]

1 Nanophotonics Technology Center (NTC), Universitat Politècnica de València, Valencia, Spain

2 Institute of optoelectronic Systems and Microtechnology (ISOM), Universidad Politécnica de Madrid, Madrid, Spain

*E-mail: vjgomher@ntc.upv.es

ABSTRACT

The catalyst-free synthesis of graphene on dielectrics prevents the damage induced by the transfer process. Although challenging, to master this synthesis would boost the integration of graphene on consumer electronics since defects hinder its optoelectronic properties. In this work, the influence of the different surface terminations of c-plane sapphire substrates on the synthesis of graphene via plasma-enhanced chemical vapour deposition (PECVD) is studied. The different terminations of the sapphire surface are controlled by a plasma etching process. A design of experiments (DoE) procedure was carried out to evaluate the major effects governing the etching process of four different parameters: i.e. discharge power, time, pressure and gas employed.

In the characterization of the substrate, two sapphire surface terminations were identified and characterized by means of contact angle measurements, being a hydrophilic (hydrophobic) surface the fingerprint of an Al-(OH-) terminated surface, respectively. The defects within the synthesized graphene were analysed by Raman spectroscopy. Notably, we found that the $I_D/I_G$ ratio decreases for graphene grown on OH-terminated surfaces. Furthermore, two different regimes related to the nature of graphene defects were identified and depending on the sapphire terminated surface are bound either to vacancy or boundary like defects. Finally, studying the density of defects and the crystallite area, as well as their relationship with the sapphire surface termination paves the way for increasing the crystallinity of the synthesized graphene.

I. INTRODUCTION

Graphene-based consumer electronics will not be affordable in large-scale fabrication until a reliable process of getting graphene on dielectrics without contamination and defects is attained [1], [2]. Focusing on sapphire, which is an ubiquitous substrate in optoelectronic devices, the chemical vapour deposition (CVD) technique has been exploited as a possible fabrication route to obtain graphene-on-sapphire [2] [3]. However, this technique presents several drawbacks such as the high temperature (> 900 ºC) needed for the precursor to dissociate and nucleate the graphene on the non-catalytic surface [4] [5] [6]. On the other hand, plasma enhanced CVD (PECVD) arises as a possible technological solution for reducing the demanding temperatures by aiding the process with the ignition of a plasma [7] [8]. The plasma promotes the decomposition of hydrocarbons, generating a large density of active radicals and species which react with the substrate surface, significantly increasing the nucleation rate of graphene. The competition between the nucleation and growth of graphene or its etching governs the PECVD synthesis of graphene. Either the growth or the etching state are determined by the process parameters, e.g. discharge power, temperature, hydrogen concentration or pressure [9] [10].

The properties of sapphire have been extensively studied for its use as substrate in the synthesis of III-V semiconductors. There are different surface reconstructions and three stable terminations for c-plane sapphire which strongly influence the properties of its surface [11]. For example, the surface termination determines the properties of III-V semiconductors grown on the sapphire surface [12] [13] [14]. Theoretical calculations have reported that the single Al-termination has the lowest surface energy whereas the oxygen termination displays the largest [15] [16] [17]. On the other hand, the O-termination is stable only when hydrogen is present in the surface [18] [19]. Upon heat treatment the surface structure is reversible [20] [21].

Considering the case of graphene synthesis on sapphire, several works report different nucleation behaviours depending on the crystal orientation or surface reconstruction of sapphire. For example, because of the catalytic behaviour of its surface, the nucleation density is largely increased when the synthesis is carried out on r-plane sapphire [22] [23]. On the other hand, Mishra et al. reported a significant increase in the measured mobility of graphene synthesized by CVD on a c-plane sapphire including a $H_2$ thermal etching at 1180 ºC [1]. They identified the oxygen deficient ($\sqrt{31} \times \sqrt{31}$) R±9° reconstruction of sapphire via low energy electron diffraction (LEED) measurements and noted that the catalytic behavior of surface Al atoms increases the quality of their synthesized graphene, as compared with the pristine c-plane sapphire. However, to the best of our knowledge, the influence of the different surface terminations of sapphire on the PECVD synthesis of graphene has not yet been addressed.

In this work, the effect of the different sapphire surface terminations on graphene grown by PECVD is reported. Moreover, the sapphire surface terminations are controlled within the synthesis chamber by means of a plasma etching process, whose major parameters governing the process are the plasma power and its gas chemistry [24] [25]. The induced modifications on the sapphire substrate have been characterized by means of contact angle and atomic force microscopy (AFM) measurements. Afterwards, a defect analysis of the synthesized graphene was carried out using the Raman spectroscopy technique. Two different regimes of the defects are found (either vacancy or boundary like defects), whereas the study of the crystallinity and the density of defects within the synthesized graphene yield a valuable insight into promoting larger size of the graphene grains on the sapphire substrate.

## II. Materials and methods

### A. Substrate cleaning

Pristine 4-inch c-plane sapphire wafers (PHOTONEXPORT Epi-Ready), covered with a protective PMMA resist to avoid splinters, were diced in 10 x 10 mm squares. The diced sapphire substrates were subjected to the following cleaning procedure: First, the PMMA resist was removed with an $O_2$ plasma for 600 seconds at 400 W and 1.5 mbar (PVA TEPLA 200, PVA TePla AG). Afterwards, the following two-solvent method was employed: to remove any particles they were rinsed below running deionized water for 30 seconds, blown dry in $N_2$ and immersed in acetone for 300 seconds at room temperature (RT). Finally, the sapphire substrates were sonicated in isopropyl alcohol (IPA) during 300 s at RT, blown dry with $N_2$ and introduced into the PECVD system (BlackMagic 6-inch, Aixtron Nanoinstruments Ltd.).

*B. Sapphire etching: Design of experiments*

First, the top and bottom heaters of the synthesis chamber were heated up to 900 ºC and 800 ºC respectively, with a heating rate of 200 ºC/min in a pure Ar atmosphere at 4 mbar. To eliminate undesired temperature gradients within the synthesis chamber, the heaters remained at the target temperatures for 600 seconds at 800 ºC. Afterwards, the sapphire etching process was initiated. The series of experiments were decided based on the table of signs or $L_8$ orthogonal array (Table 1) to study the influence of four factors using a two-level fractional factorial design $2^{4-1}$ [26]. The fractional factorial design, implemented with the Statgraphics centurion XVIII software, screens major influences of the different factors under study along with the interactions between them. The four selected factors were: discharge power, process pressure, gas and process time. In our case, the process temperature was set to 800 ºC after optimizing the synthesis of graphene on pristine c-plane sapphire surfaces (see supplementary material). The substrate was then cooled down at a rate of 15 ºC/min and the samples unloaded at temperatures below 200 ºC. Afterwards the sapphire substrates were characterised by means of water contact angle measurements in a ramé-hart automatized goniometer and atomic force microscopy (AFM) measurements (MultiMode 8-HR, Bruker) in tapping mode.

*Table 1 Table of sings employed for the $2^{4-1}$ fractional factorial design. Signs are in brackets as a guide to understanding the creation of the $L_8$ orthogonal array*

| | Factors | | | |
|---|---|---|---|---|
| | a | b | c | abc |
| **Run Number** | **Pressure [mbar]** | **Power [W]** | **Time [s]** | **Gas type** |
| 1 | 4 (−) | 100 (+) | 600 (+) | Ar (−) |
| 2 | 6 (+) | 100 (+) | 600 (+) | $N_2$ (+) |
| 3 | 6 (+) | 50 (−) | 300 (−) | $N_2$ (+) |
| 4 | 4 (−) | 50 (−) | 300 (−) | Ar (−) |
| 5 | 6 (+) | 100 (+) | 300 (−) | Ar (−) |
| 6 | 4 (−) | 50 (−) | 600 (+) | $N_2$ (+) |
| 7 | 6 (+) | 50 (−) | 600 (+) | Ar (−) |
| 8 | 4 (−) | 100 (+) | 300 (−) | $N_2$ (+) |

## C. PECVD growth of graphene.

For minimizing variables induced by the PECVD system, the graphene synthesis was carried out within the same run in every sapphire sample. Additionally, a pristine c-plane sapphire substrate was introduced (reference sample). The samples were cleaned using the two-solvent method described above and introduced into the PECVD chamber. Similar to the heating process explained before, the system stayed for 600 s in a pure Ar atmosphere with a process pressure of 4 mbar once the top and bottom heaters reached 900 °C and 800 °C after a heating rate of 200 °C/min. Afterwards, the supply of Ar was halted while the reactive gas (methane, $CH_4$) was introduced with the following gas ratio 1:5:5; $CH_4$, $N_2$ and $H_2$ respectively and the DC plasma turned on with a discharge power of 100 W. The growing process was set to 1200 s. The synthesis conditions were optimized for the PECVD growth of graphene on pristine sapphire (see supplementary material). The substrate was then cooled at a rate of 15 °C/min and the samples unloaded at temperatures below 200 °C.

*Table 2 Process parameters for the synthesis of graphene on the etched sapphire substrate*

| Temperature [°C] | Pressure [mbar] | Plasma power [W] | $N_2$ [sccm] | Ar [sccm] | CH4 [sccm] | $H_2$ [sccm] | Time [s] |
|---|---|---|---|---|---|---|---|
| 800 | 4 | 100 | 100 | 0 | 20 | 300 | 1200 |

*D. Raman spectroscopy characterization*

The structural quality of the synthesized graphene was characterized by Raman spectroscopy at room temperature. A confocal Raman imaging microscope (alpha 300R, WITec) was employed in the backscattering configuration using a 100x objective and a 600 gr/mm grating with 2.8 cm$^{-1}$ resolution. The excitation energy (wavelength) from the laser diode module was 2.33 eV (532 nm) and the power set to 25 mW. In order to have a representative measure of the graphene on the sapphire surface, each sample was characterized by means of 40 Raman spectra (2 accumulations, 12 s integration time) at different locations using an x-y piezo-scanner stage. The Raman spectra were then fitted using a Lorentzian function (Figure S1), extracting information of peak position, full width at half maximum (FWHM), intensity and area.

*E. Contact angle measurements*

The needle-in-sessile-drop method was used to measure the advancing and receding contact angle of the etched sapphire substrates before and after the graphene synthesis using an automatized goniometer (90-U3-PRO, ramé-hart instrument co.) [27]. The measurements were taken at room temperature with no humidity control. Before each measurement, the samples were cleaned by the two-solvent method described previously to avoid experimental errors caused by dirt and impurities. First, a deionized water droplet of approximately 1 ml was pumped out by a motorized micro syringe normal to the sample surface. Water was then added to the drop at very low volumes (1 µl). The contact angle in each volume step was then recorded to measure the advancing contact angle. To obtain the receding contact angle, the water of the droplet was pumped in (1 µl) until the minimum angle was measured. This advancing-receding iteration was performed for 5 times, disregarding the first maximum and minimum values for the statistics because they are influenced by external factors such as airborne contamination.

# III. RESULTS

## A. *Results of Contact angle measurements*

The density of hydroxyl groups, which strongly adsorb water molecules, determines the hydrophilicity of the oxide surfaces [28]. In the case of c-plane sapphire, the Al-termination, which acts as a strong Lewis acid, promotes the H2O adsorption [19] [29]. Because of the high reactivity of the aluminum termination in water the surface is easily hydroxylated with free hydroxyl groups and thus the surface becomes more hydrophilic, as observed in the improved wettability of the samples processed by N2 based etching (Figure 1). In these etching processes, the largest reduction (45 %) of the contact angle (36°) is measured for the combination of 6 mbar, 100 W and 600 s as compared with the reference sample (66°).

However, the sapphire surface can also exhibit terminal oxygen bridges ($Al_2O^-$ species) which, if sufficient hydrogen is supplied, become terminal hydroxide bridges ($Al_2(OH)$ species) [19]. When the first layer of water molecules interacts with this surface, few H atoms of water tend to form hydrogen bonds with the oxygen atoms within the terminal hydroxide bridges of the surface. Thus, a physical barrier is created which prevents the hydroxylation of the aluminum atoms, preserving the Lewis acid sites. This process is the reason behind the increased hydrophobicity of the surface, as observed in the c-plane sapphire substrates subjected to an Ar based etching (except for the combination of 4 mbar, 50 W and 300 s) [29] [30]. This particular combination reduces the contact angle of sapphire (50°). On the other hand, the combination of 4 mbar, 100 W, 600 s and Ar; reports the largest increment (136 %) of contact angle (90°) whereas the contact angles measured for the sapphire substrate etched with the combinations of 6 mbar, 100 W and 300 s and 6 mbar, 50 W, 600 s are 80° and 70° respectively. In the following, the OH-terminated sapphire substrates are considered those that increase the hydrophobicity as compared to the reference sample.

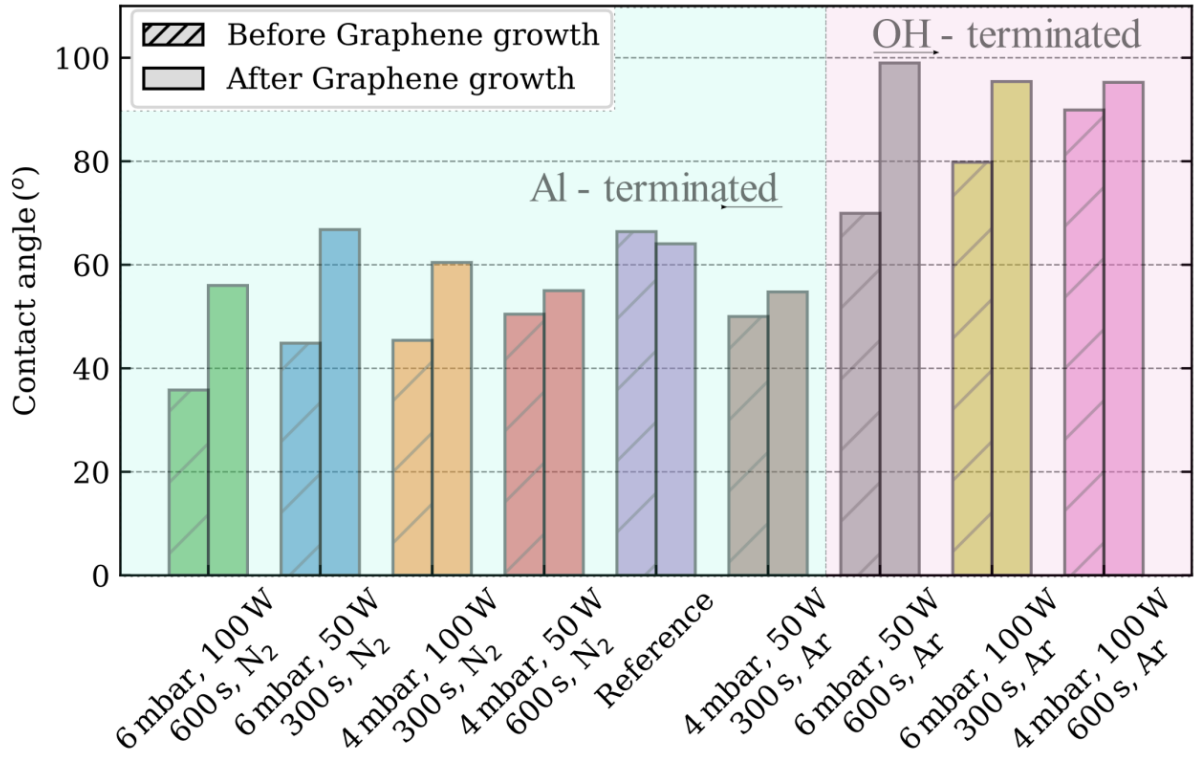

*Figure 1 Advancing contact angle measurements of the c-plane sapphire substrates based on the etching recipes*

Furthermore, the wettability properties of sapphire can be modified to a large extent using patterned nanostructures [31]. However, this effect is disregarded in our experiments, as the AFM measurements do not provide large variations of roughness after the etching processes (see supplementary material).

When the contact angle of graphene is evaluated, a large controversy exists because of its "wetting transparency", as the graphene layer modifies the adsorption energy between the water molecule and the substrate [32]. Contact angle measurements on free-standing graphene have shown the hydrophilic properties of graphene [33]. However, in our experiments, when the graphene is grown on the sapphire substrate, the hydrophobicity increases towards the contact angle of graphite (~90º) independently of the etching recipe.

*B. Results of the DOE*

The study of main effects e.g. plasma power or process time on the parameter under study e.g. contact angle allows us to find the influence of an etching factor [26]. As it has been already observed, the influence of the gas employed is remarkable thus it is reasonable to analyse the main effects either for the $N_2$ or Ar etching processes (Figure 2). When the $N_2$ gas is employed, the sapphire surface is modified and acts more hydrophilic which indicates an Al-terminated sapphire surface [19] [29]. When evaluating the main effects influencing the contact angle of the etched sapphire surface: discharge power and pressure have a larger influence than for example the etching time, since the steeper the slope, the larger is the influence of that particular parameter.

However, when Ar is employed the wettability of sapphire is completely different and the surface becomes more hydrophobic, which is indicative of an OH-termination of the sapphire surface [19]. In this case, our analysis shows that the pressure has a lower influence on the contact angle than the discharge power or the time of the etching process.

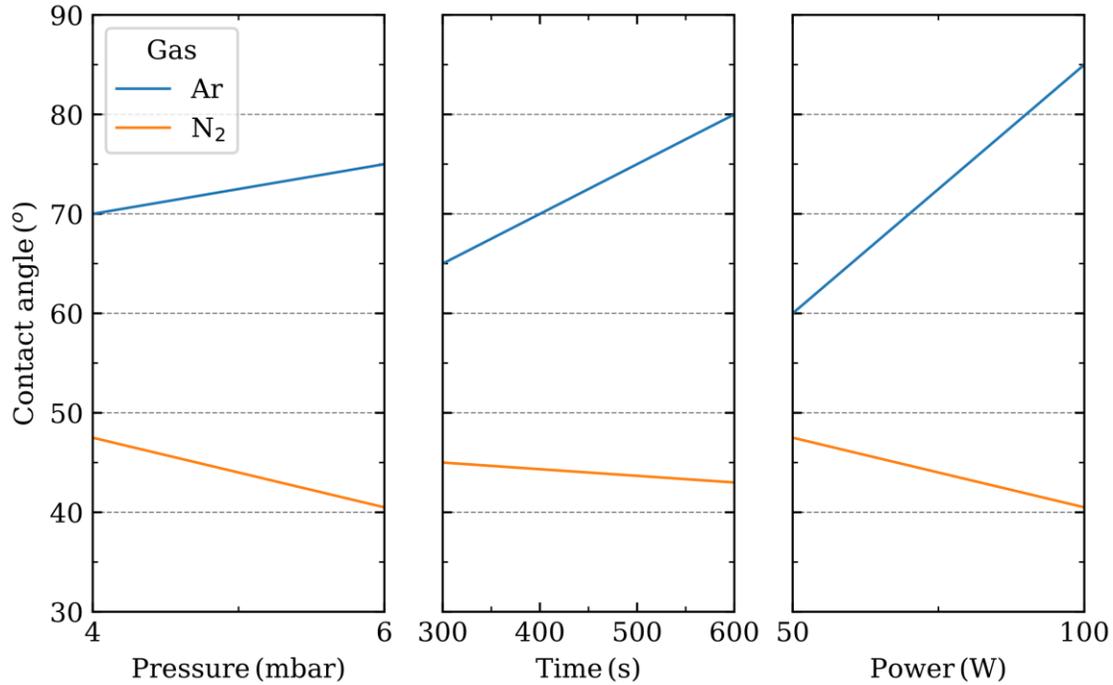

*Figure 2 Main effects plot of pressure, power and time depending on the gas employed during the sapphire etching process. A: Etching process carried out in a $N_2$ atmosphere. B: Etching process carried out in an Ar atmosphere*

When the analysis of the main effects evaluates the roughness of the sapphire substrate after the etching process (Figure S7), the parameters with a larger influence are, as expected, the etching time and the discharge power (Figure 3). The former, together with the minor influence of the employed gas (i.e. $N_2$ or Ar) indicates that the etching process is carried out at a similar etching rate and that no reactive radicals are governing the process [34]. The discharge power provides the potential difference to accelerate the ions towards the sapphire substrate, thus transferring their kinetic energy to the surface atoms. On the other hand, the mean-free-path between collisions within the gas phase is determined by the process pressure. At larger pressures the ions get thermalized as they experience more collisions in their travel towards the sapphire substrate. As observed from the main effect analysis, the surface roughness of the sapphire surface is slightly modified by the directionality of the impinging ions towards the surface.

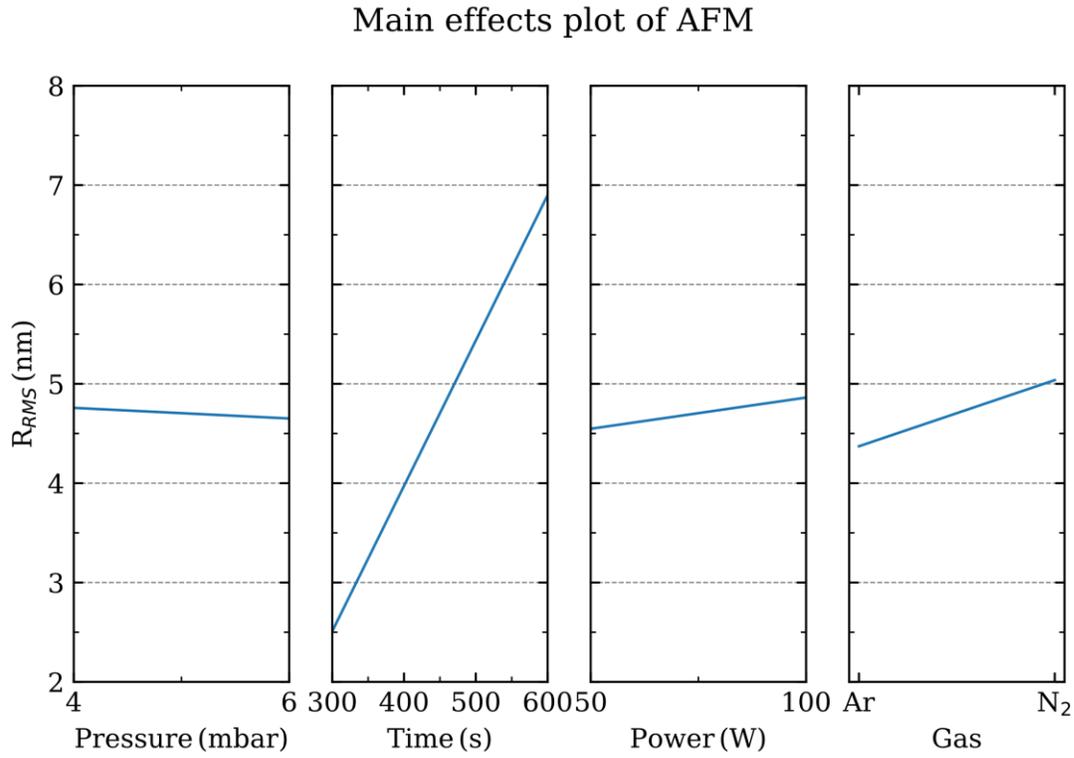

*Figure 3 Main effects plot of the root mean square roughness ($R_{RMS}$) for the pressure, power, time and gas employed during the sapphire etching process*

## C. Raman spectroscopy

The Raman fingerprint modes of pristine graphene are the G band at around 1583 cm$^{-1}$ together with the defect-related bands: the D (1300 cm$^{-1}$), its overtone the 2D (2680 cm$^{-1}$) and the D' (1620 cm$^{-1}$) band[35]. Although the 2D band is usually referred as the overtone of the D band, it is the most prominent feature in graphene since no defects are required for the activation of second order phonons [36]. When the Raman spectra of the synthesized graphene are evaluated, the characteristic graphene bands fade in several Raman spectra depending on the sapphire etching process carried out. Significantly, the first relationship between the induced terminations within the etched c-plane sapphire surfaces and the synthesis of graphene arises when evaluating the coverage of the sapphire surface. For example, the full-coverage is not completely achieved when the synthesis is carried out in the Al-terminated (more hydrophilic) sapphire surfaces. This, the full coverage, is studied via the data dispersion shown by the histograms of the fitted full-width-at-half-maximum (FWHM) values for the graphene D, G and 2D bands (see supplementary material: Table S3). Among the Al-terminated sapphire surfaces the full coverage is achieved for the reference sample along with one $N_2$ based etching process (4 mbar, 50 W and 600 s) and, independently of the sapphire surface termination, every Ar based etching process.

When graphene is grown on sapphire with an OH-termination, the graphene D and 2D bands experience (Figure 4) a red shift above ~10 cm$^{-1}$ whereas the G and D' peaks experience a ~10 cm$^{-1}$ blue shift as compared with the reference sample. On the other hand, when graphene is grown on etch induced Al-terminated sapphire surfaces there are no remarkable shifts of the D, G, D', and 2D band positions.

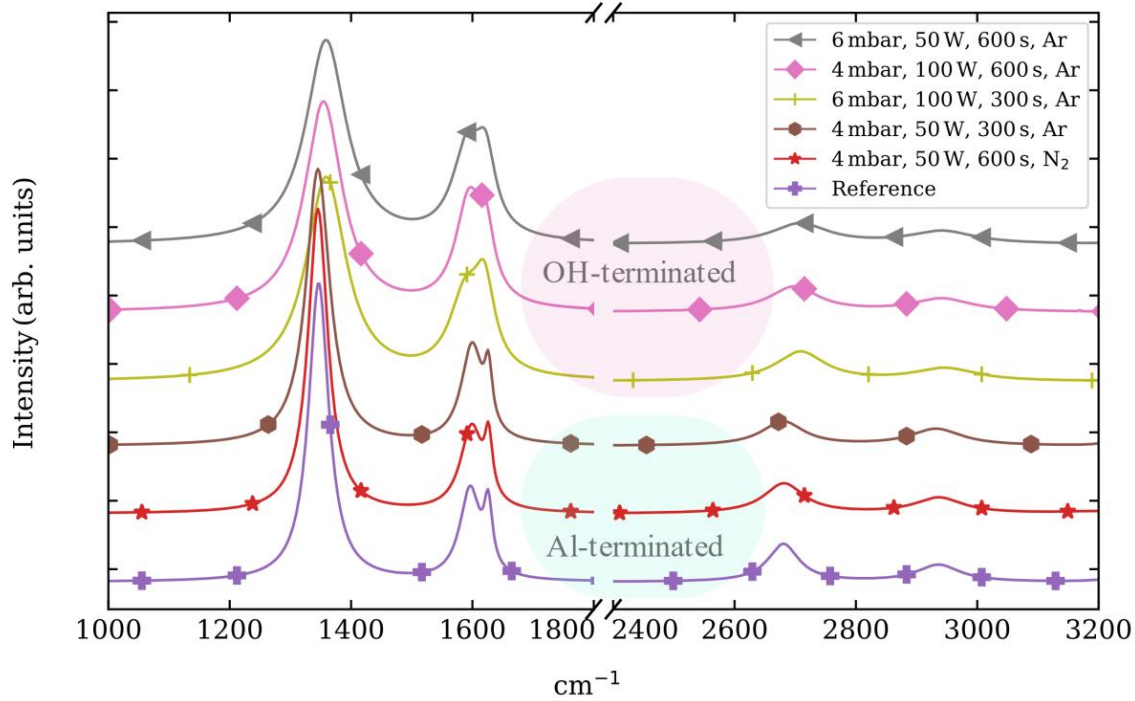

*Figure 4 Standard normal variate normalization of the graphene Raman spectra grown by PECVD on sapphire substrates etched with different combinations of pressure, power, time and gas [37]. The lines are shifted in intensity for the aid of visualization*

The $I_D/I_G$ ratio is usually employed to provide a measure of defects in graphene (Figure 5A). Within the Al-terminated sapphire surfaces the $I_D/I_G$ ratios are comparable to those of the reference sample (3.35). However, the $I_D/I_G$ ratio decreases for graphene grown on OH-terminated sapphire surfaces where a reduction of 36 % is observed (Ar, 4 mbar, 100 W and 600 s).

On the other hand, the structural quality is usually evaluated by the $I_{2D}/I_G$ ratio. Independently of the either Al- or OH- terminated sapphire surface, the $I_{2D}/I_G$ ratio decreases as compared with the reference sample ($I_{2D}/I_G$ ratio ~0.42). Both, smaller (9 %) and larger (42 %) reductions are observed for the graphene grown on OH-terminated sapphire surfaces. Interestingly, these boundaries correspond to the smaller and larger variation of surface roughness $R_{RMS}$ (Figure S7). However, this relationship is constrained to the OH-terminated surfaces, since for the Al-terminated sapphire surfaces the larger decrease (27 %) occurs to the sample with smaller $R_{RMS}$.

The presence of sharp defect related D and D' bands, together with the presence of the D+D' band and the position of the G band allows us to identify the synthesized graphene within the stage I of the amorphization trajectory in graphite [38]. In this stage, the D-band scattering is proportional to the average distance between nearest defects ($L_D$) or the defect density ($\sigma$), thus $I_D/I_G \propto 1/L_D^2 \propto \sigma$ which can be calculated, in terms of the laser excitation wavelength $\lambda_L$ (nm) after Equation 1 and Equation 2 respectively [39].

| | |
|---|---|
| $$L_D^2 \ (nm^2) = 1.8 \times 10^{-9} \lambda_L^4 \left(\frac{I_D}{I_G}\right)^{-1}$$ | Equation 1 |
| $$\sigma \ (nm^{-2}) = 7.3 \times 10^{-9} \lambda_L^4 \left(\frac{I_D}{I_G}\right)$$ | Equation 2 |

Therefore the $I_D/I_{D'}$ ratio does not depend on the defect concentration but on the nature of the defects [40]. This ratio has been associated to boundary-like defects for values close to 3.5 and to vacancy-like defects when $I_D/I_{D'}$ tends to 7. Within our samples there are two different regimes depending on the plasma induced sapphire surface terminations (Figure 5 B). The graphene samples grown on the Al-terminated sapphire surfaces reports an $I_D/I_{D'}$ ratio which tends towards the vacancy-like regime. On the other hand, the OH-terminated surfaces tend to the boundary-like regime.

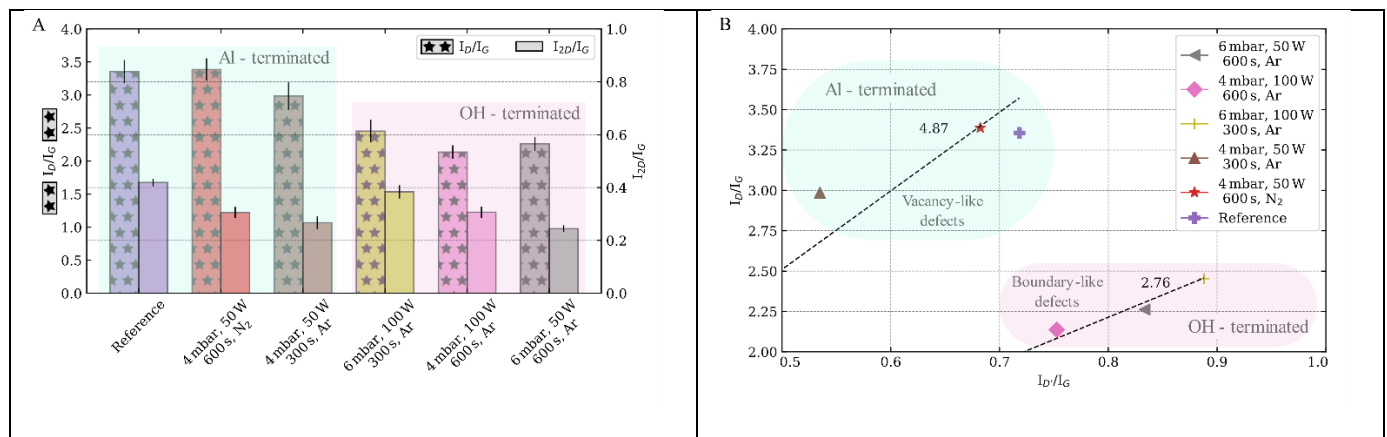

*Figure 5 Intensity ratios of the characteristic graphene Raman bands of the samples synthesized on sapphire substrates etched with different combinations of pressure, power, time and gas. A: $I_D/I_G$ and $I_{2D}/I_G$ ratios. B: $I_D/I_G$ vs. $I_{D'}/I_G$ ratios which can be employed to evaluate the nature of defects.*

Furthermore, the contributions of point and line defects in the Raman spectra can be further investigated from the $I_D/I_G$ ratio [41]. The point defects are considered as 0D defects, and they are characterized by the average distance between nearest defects ($L_D$). For a perfect graphene $L_D \to \infty$, whereas for a fully disordered graphene $L_D \to 0$. On the other hand, line defects (1D) are evaluated by the crystallite size ($L_a$) (Equation 3) or by the crystallite area ($L_a^2$) [42].

$$L_a \ (nm) = (2.4 \times 10^{-10}) \lambda_L^4 \left(\frac{I_D}{I_G}\right)^{-1}$$

Equation 3

Similar to the analysis method proposed by Cançado et.al. [41], the effect of the plasma induced surface terminations is evidenced in the variations of the defect density σ and crystallite area $L_a^2$ (Figure 6). Interestingly, the Al-terminated sapphire surfaces correspond to a larger number of defects and smaller crystallite area. On the other hand, OH-terminated surfaces tend to have smaller defect densities and larger crystallite areas. Despite a trend is observed, one sample of graphene grown on an Al-terminated sapphire surface reports similar defect density and crystallite area as those with OH-sapphire surfaces. Particularly, this Al-terminated sapphire surface, etched with an Ar based process, reports the smaller $R_{RMS}$ roughness among the etched sapphire substrates (Figure S7) which could be the reason behind the crystallite area and defect density reported. However, further work should be carried out to discern the possible influence of the sapphire surface roughness on the synthesis of graphene.

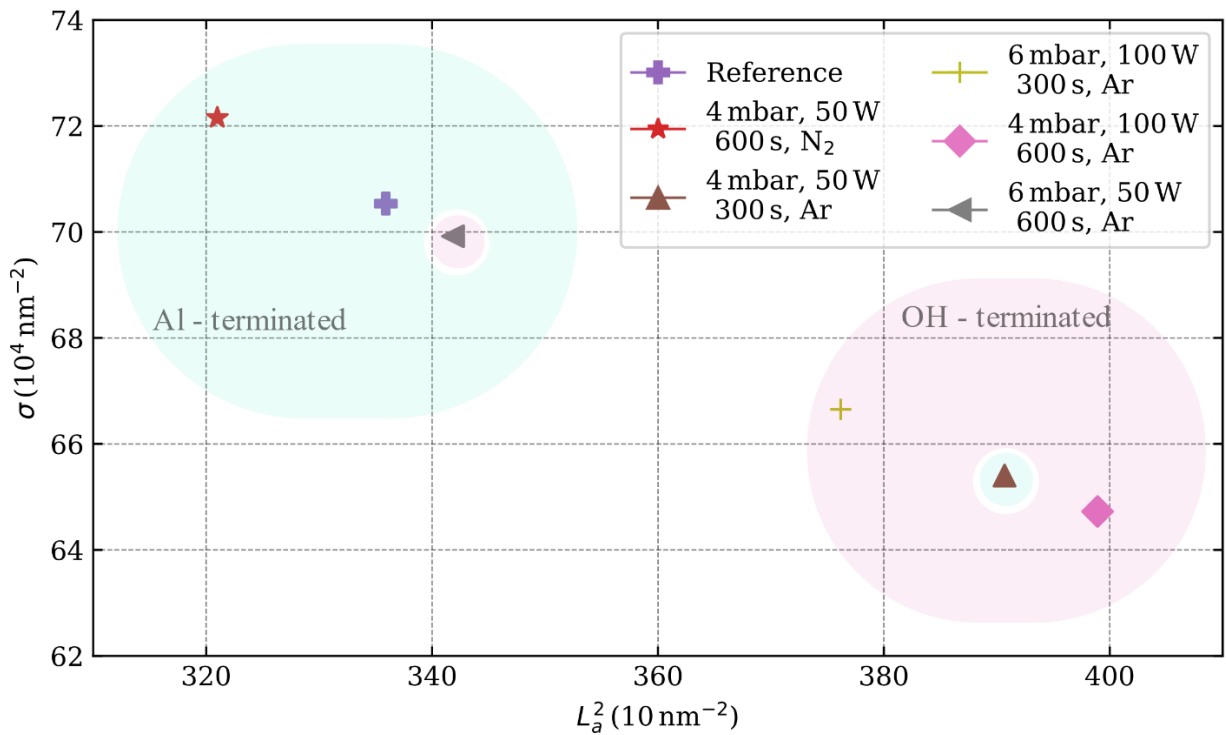

*Figure 6 Defect density ($\sigma = 1/L_D^2$) versus crystallite area ($L_a^2$) with the $L_a$ and $L_D$ values as calculated from (Equation 3) and (Equation 2) respectively.*

The commented trend of reducing the defect densities and increasing the crystallite area of graphene grown on OH-terminated sapphire surfaces is explained by the catalytic effect of the superficial aluminum. At temperatures above 500 ºC, the hydroxyls of the surface are removed, leaving highly aluminum rich surfaces thus favoring the catalytic effect of Al during the synthesis [28] [43]. There are several reports where the growth rates of graphene on different sapphire planes are evaluated and interestingly, the catalytic effect of aluminum in the surface is also reported [22] [23]. According to Ueda et al. the catalytic effect of the r-plane sapphire surface promotes the full coverage of the sapphire surface [23] [44]. However, their CVD process is carried out at 1200 ºC. In the particular case of c-plane sapphire, both Saito et al. and Ueda et al. report that the growth of graphene is only observed in the pits of Al-rich surfaces caused by the thermal desorption of oxygen atoms within the surface [22], [23], [44]. On the other hand, the synthesis route proposed in this work controls the surface termination of the c-plane sapphire using a plasma process; uses the PECVD technique which reduces the temperature needed for the precursor to dissociate thus allowing the synthesis at lower temperatures .

IV. CONCLUSION

In conclusion, our experiments show that the plasma-induced modifications of the c-plane sapphire surface play a role in the PECVD synthesis of graphene. By controlling the plasma etching process through a combination of process pressure, discharge power, time and type of gas, the sapphire surface termination can be engineered. In this work either Al- or OH- terminations of the sapphire surface have been identified by contact angle measurements. As compared with the pristine c-plane sapphire surface, the OH-terminated surface increases the hydrophobicity of the sapphire surface, whereas an Al-terminated surface reduces its contact angle with the water droplet thus being more hydrophilic. The AFM analysis demonstrate that the wettability behaviour is not due to the surface roughness created by the etching, but for a different surface chemical termination.

The quality of the PECVD synthesized graphene on the plasma etched sapphire substrates is studied by its fingerprint Raman resonances namely its D, G, D' and 2D bands. The full coverage of the sapphire surface is achieved for all OH-terminated surfaces. On the contrary, not every Al-terminated surface shows a full coverage, understood as a small dispersion of the fitted FWHM values for the D, G and 2D bands. Probably the catalytic behaviour of the highly aluminum rich surface because of the decomposition at temperatures above 500 ºC of the hydroxyls within the surface.

The $I_D/I_G$ ratio, usually employed as a measure of defectiveness in graphene, is substantially reduced when graphene is synthesized on the OH-terminated surfaces (Ar-based etching process). On the other hand, independently of the etching process, the $I_{2D}/I_G$ ratio which is usually reported as a quality measure of graphene, is reduced as compared with the pristine c-plane sapphire. The nature of graphene defects was identified by the $I_D/I_{D'}$ ratio: Al-terminated sapphire surface show vacancy-like defects in graphene whereas boundary-like defects are more prominent in the graphene grown on OH-terminated sapphire surfaces.

The representation of defect density ($\sigma = 1/L_D^2$) versus crystallite area ($L_a^2$) allows to discern how the graphene crystallite size and its defectiveness is largely influenced by the etching process carried out on the sapphire substrate. These results provide insight into the synthesis process of graphene on c-plane sapphire and ways to improve the quality of graphene without the need for changing the sapphire crystal orientation.

## V. Acknowledgments

M.S.L acknowledges financial support from the Generalitat Valenciana (Project: CIAPOS/2021/293). V.J.G acknowledges financial support from the Generalitat Valenciana (CDEIGENT/2020/009; MFA/2022/025). E.P-C. acknowledges financial support from the Generalitat Valenciana (SEJIGENT/2021/039). All authors acknowledge financial support from AGENCIA ESTATAL DE INVESTIGACIÓN of Ministerio de Ciencia e Innovacion (PID2020-118855RB-I00 and PID2021-128442NA-I00).

## VI. Declaration of competing interest

The authors declare that they have no known competing financial interests or personal relationships that could have appeared to influence the work reported in this paper.